\documentclass[aps,prl,superscriptaddress,amsfonts,amsmath,amssymb,showpacs,floatfix,reprint,longbibliography,footinbib]{revtex4-2}

\usepackage{url}
\usepackage{bm}
\usepackage{graphicx}
\usepackage{amsmath}
\usepackage{amstext}
\usepackage{amssymb}
\usepackage{amsfonts}
\usepackage{amsbsy}
\usepackage{verbatim}
\usepackage{color}
\usepackage[colorlinks=true, urlcolor=blue, linkcolor=blue, citecolor=blue, pdftex]{hyperref}
\usepackage[capitalise]{cleveref}
\usepackage{multirow}
\usepackage{floatrow}
\usepackage{float}
\usepackage{tikz}
\usepackage{gensymb}
\usepackage{textcomp}
\usepackage{enumitem}
\usepackage[version=3]{mhchem}
\usepackage{afterpage}
\usepackage{pbox}
\usepackage{makecell}
\usepackage{dsfont}
\usepackage{siunitx}
\usepackage{fancyhdr}
\usepackage{stackengine,wasysym,scalerel}
\usepackage{latexsym}
\usepackage{cancel}
\usepackage{array, multirow, bigdelim, makecell, booktabs}
\usepackage[misc]{ifsym}

\newcommand{\bmq}{\mathbf{q}}

\begin{document}

\title{Quantum effects on unconventional pinch point singularities}
  
  \author{Nils Niggemann}
\email{nils.niggemann@fu-berlin.de}
\affiliation{Dahlem Center for Complex Quantum Systems and Fachbereich Physik, Freie Universit{\"a}t Berlin, D-14195 Berlin, Germany}
\affiliation{Helmholtz-Zentrum Berlin f{\"u}r Materialien und Energie, D-14109 Berlin, Germany}
\affiliation{Department of Physics and Quantum Center for Diamond and Emergent Materials (QuCenDiEM), Indian Institute of Technology Madras, Chennai 600036, India}
\author{Yasir Iqbal}
\affiliation{Department of Physics and Quantum Center for Diamond and Emergent Materials (QuCenDiEM), Indian Institute of Technology Madras, Chennai 600036, India}
\author{Johannes Reuther}
\affiliation{Dahlem Center for Complex Quantum Systems and Fachbereich Physik, Freie Universit{\"a}t Berlin, D-14195 Berlin, Germany}
\affiliation{Helmholtz-Zentrum Berlin f{\"u}r Materialien und Energie, D-14109 Berlin, Germany}
\affiliation{Department of Physics and Quantum Center for Diamond and Emergent Materials (QuCenDiEM), Indian Institute of Technology Madras, Chennai 600036, India}

  \date{\today}

\begin{abstract}
Fracton phases are a particularly exotic type of quantum spin liquids where the elementary quasiparticles are intrinsically immobile. These phases may be described by unconventional gauge theories known as tensor or multipolar gauge theories, characteristic for so-called type-I or type-II fracton phases, respectively. Both variants have been associated with distinctive singular patterns in the spin structure factor, such as multifold pinch points for type-I and quadratic pinch points for type-II fracton phases. Here, we assess the impact of quantum fluctuations on these patterns by numerically investigating the spin $S=1/2$ quantum version of a classical spin model on the octahedral lattice featuring exact realizations of multifold and quadratic pinch points, as well as an unusual pinch line singularity. Based on large scale pseudo fermion and pseudo Majorana functional renormalization group calculations, we take the intactness of these spectroscopic signatures as a measure for the stability of the corresponding fracton phases. We find that in all three cases, quantum fluctuations significantly modify the shape of pinch points or lines by smearing them out and shifting signal away from the singularities in contrast to effects of pure thermal fluctuations. This indicates possible fragility of these phases and allows us to identify characteristic fingerprints of their remnants.
\end{abstract}

\maketitle

{\it Introduction.---} 
A particularly fascinating physical situation arises when a system of interacting spins realizes an emergent gauge theory, which is one of the defining properties of a quantum spin liquid~\cite{Savary-2017}.
Various different types of gauge theories may be realized in such phases.
For example, quantum spin ice represents a variant of a quantum spin liquid, where an emergent U(1) gauge theory on a pyrochlore lattice establishes an astonishing analogy to three-dimensional electromagnetism including emergent photons and an effective fine-structure constant \cite{Hermele2004,Pace2021}. The key ingredient enabling these non-trivial properties is the gauge constraint which, in the charge-free sector of a U(1) gauge theory, takes the form of a Gauss law ${\bm \nabla}\cdot{\bm E}(\bm{r})=0$.

Meanwhile, generalizations of the standard U(1) gauge theories have become a new focus of theoretical investigations where the vector form of the Gauss-law is replaced by a tensor structure~\cite{Xu-2006,Pretko2017,Prem2018,Yan2020}, e.g. $\sum_{\mu\nu}\partial_\mu\partial_\nu E_{\mu\nu}(\bm{r})=0$, known as tensor gauge theories describing so-called fracton spin liquids~\cite{Nandkishore-2019,Pretko-2020}. The most remarkable consequence of this generalization is that, besides the effective charge of a quasiparticle, multipole moments of charges become conserved quantities giving rise to excitations with fractionalized mobility~\cite{Pretko2017_1}. Two cases can be distinguished~\cite{Vijay-2016}: In type-I fracton phases~\cite{Chamon-2005,Bravyi-2011,Vijay-2015}, described by symmetric tensor gauge theories, the quasiparticles are either completely immobile or have a residual mobility along subdimensional manifolds. Otherwise, in type-II fracton phases~\cite{Haah-2011,Yoshida-2013,Castelnovo-2012} all quasiparticles are completely immobile. In the associated multipolar gauge theories the Gauss law contains derivatives of different orders restricting charge configurations to certain fractal patterns~\cite{Bulmash-2018,Schmitz-2019,Gromov-2019,Gromov-2020}. Remarkably, fracton phases also attract interest in fields such as quantum information~\cite{Terhal-2015,Schmitz-2018} and high energy physics~\cite{yan19,yan192,Seiberg-2020,Gorantla-2021}.

Recently, important steps have been undertaken to bring the rather abstract theoretical research on fracton phases closer to the established field of quantum magnetism and to experiments. For example, it has been found that 
type-I fracton phases manifest themselves in multifold pinch-points~\cite{Prem2018} in the spin structure factor [Fig.~\ref{fig:Multifold}(a)], generalizing the famous twofold pinch points known from conventional U(1) spin liquids [Fig.~\ref{fig:octochlore}(c)]. Likewise, type-II fracton phases have been argued to be associated with quadratic pinch points [Fig.~\ref{fig:Parabolic}(d)] where contour lines exhibit a characteristic parabolic shape~\cite{Hart2022}. On a different front, a class of simple classical spin models have been identified \cite{Benton2021} 
which give straightforward access to classical spin liquids described by tensor gauge theories and to unconventional pinch points in the spin structure factor.
However, it is an open but experimentally relevant question how stable these phases are under modification from the ideal situations in which they are defined, e.g., by allowing for quantum fluctuations.

In this letter, we study the effects of quantum fluctuations on the ground state and finite-temperature phases of the classical spin model in Ref.~\cite{Benton2021}-- the so-called octochlore model -- whose three dimensional octahedral lattice is
realized in rare-earth antiperovskites ~\cite{Henley2013,Szabo2022}.
This model represents a showcase example for exotic classical spin liquids:
Apart from known twofold and multifold pinch points we identify exact realizations of quadratic pinch points~\cite{Hart2022} as well as unconventional pinch line singularities~\cite{Benton2016}. We add quantum fluctuations to the system by promoting it from a classical ($S\to\infty$) to a quantum $S=1/2$ Heisenberg model which is then numerically treated via two powerful quantum many-body techniques, the pseudo fermion and the pseudo Majorana functional renormalization group. Overall, we find that exotic pinch point features are drastically affected by quantum fluctuations and appear more fragile compared to conventional twofold pinch points.

\begin{figure}
    \centering
    \includegraphics[width = \linewidth]{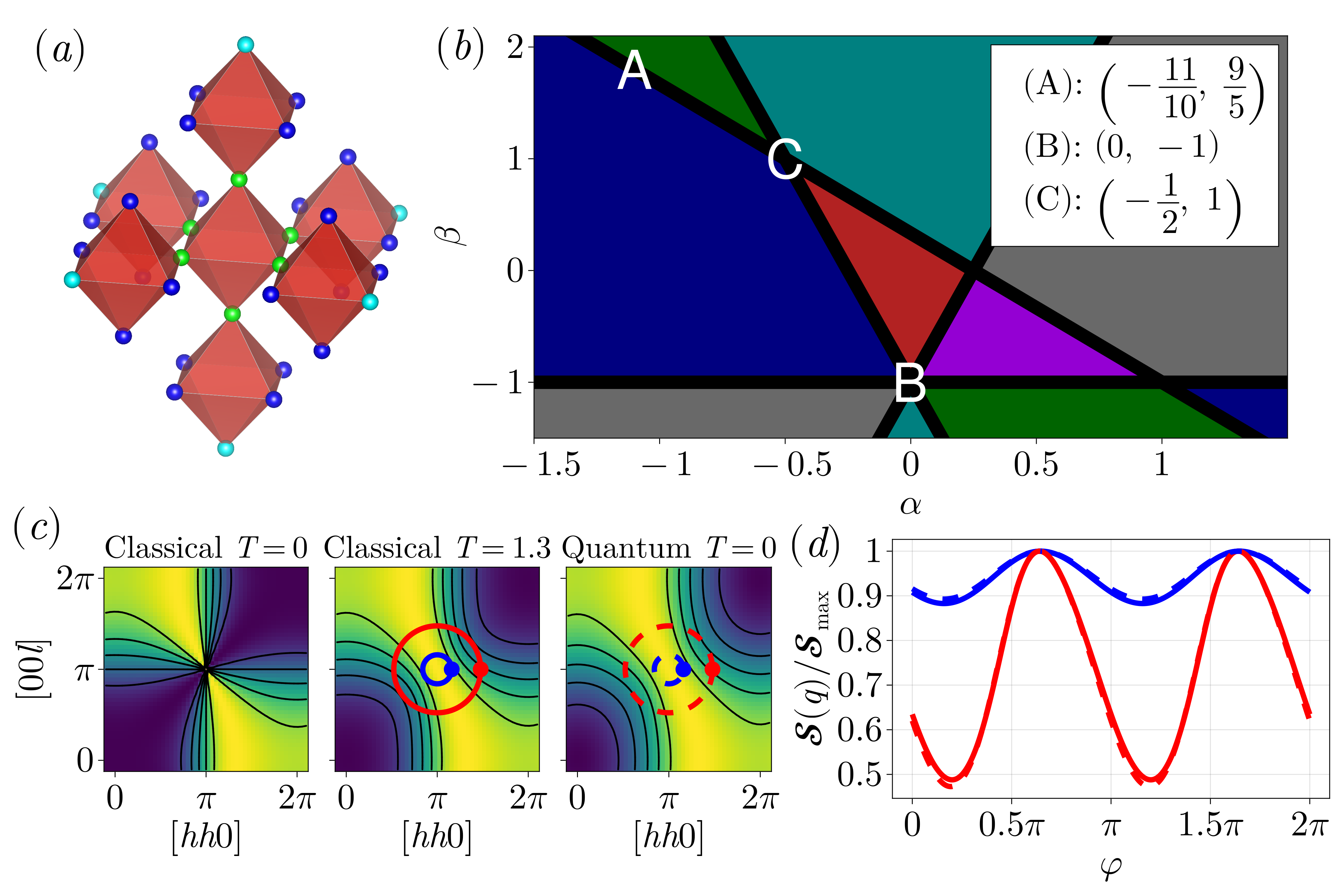}
    \caption{
(a) Octochlore model: Differently weighted sites in Eq.~(\ref{eq:OctochloreH2}) are indicated by different colors. (b) Phase diagram of the model from Ref.~\cite{Benton2021}. The labels A, B, C [with parameters $(\alpha,\beta)$ given in the inset] indicate the locations of multifold pinch points, quadratic pinch points and pinch lines as shown in \cref{fig:Multifold,fig:Pinchline,fig:Parabolic}, respectively. (c) Spin structure factor of a twofold pinch point at $\alpha = \beta = 0$ for the classical and quantum model. (d) Spin structure factor $\mathcal{S}(\bmq)$ along circular paths indicated in (c) normalized to their maxima.}
    \label{fig:octochlore}
\end{figure}

{\it Unconventional gauge theories from an octochlore model.---}
The octahedral lattice consists of corner-sharing octahedra and is defined by simple cubic lattice vectors $\mathbf{a}_m \in \left\{ (1,0,0), (0,1,0), (0,0,1) \right\}$ together with a three site basis $\mathbf{b}_m =  \mathbf{a}_m/2$.
The Hamiltonian of the octochlore model~\cite{Benton2021} is constructed as the sum of squared vectors $\mathbf{M}_{\textrm{oct}, \alpha \beta}$ over all elementary octahedra
\begin{align}
    H &= \frac{J}{2} \sum_\textrm{oct}{\mathbf{M}^2_{\textrm{oct}, \alpha \beta}}, \label{eq:OctochloreH} 
\end{align}
where $\mathbf{M}_{\textrm{oct}, \alpha \beta}$ is the sum of spins in a cluster, weighted by dimensionless parameters $\alpha$, $\beta$,
\begin{align}
    \mathbf{M}_{\textrm{oct}, \alpha \beta} &= \sum_{i \in \textrm{oct}} \mathbf{S}_i + \alpha \sum_{i \in \langle \textrm{oct} \rangle } \mathbf{S}_i + \beta \sum_{i \in \langle \langle \textrm{oct} \rangle\rangle }\mathbf{S}_i.\label{eq:OctochloreH2}
\end{align}
Here, a reference octahedron ``$\textrm{oct}$'' is given by the green sites in \cref{fig:octochlore}(a), while its closest surrounding sites ``$\langle \textrm{oct} \rangle $'' and further distant sites ``$\langle \langle \textrm{oct} \rangle\rangle $'' are colored blue and cyan, respectively. Henceforth, we set the energy scale such that the maximal Heisenberg coupling between two spins is equal to one.

For classical spins $\mathbf{S}_i$, the system's extensively degenerate ground states follow from the constraints
$\mathbf{M}_{\textrm{oct},\alpha \beta} = 0$ which constitute discrete versions of Gauss's law. These constraints can be expressed in reciprocal space as $\sum_m L_m(\bmq)\mathbf{S}_m(\bmq) = 0$~\cite{Benton2021}, where $m =1,2,3$ label the sublattices, $\mathbf{S}_m(\mathbf{q})$ is the Fourier-transformed spin on sublattice $m$ and $L_m(\bmq)$ is the $m$-th component of the so-called constraint vector. Normalized constraint vectors $\tilde{L}_m(\bmq)=L_m(\bmq)/\sqrt{\sum_n (L_n(\bm q))^2}$ can be defined over the entire $\bmq$ space except at singular points $\bmq^\star$ where $L_m(\bmq^\star)=0$ for all $m$. For isolated points $\bmq^\star$ in momentum space and with $\tilde{L}_m(\bmq)$ defined on the unit sphere $S^2$ one can assign a topological index to the defect configuration $\tilde{L}_m(\bmq)$ around $\bmq^\star$ defined by the second homotopy group of $S^2$, which is the Skyrmion number $Q$~\cite{Mermin-1979,Kleman}. As demonstrated in Ref.~\cite{Benton2021} non-trivial $Q\neq0$ give rise to pinch points at $\bmq=\bmq^\star$ in the equal-time spin structure factor $\mathcal{S}(\bmq) \equiv \langle \mathbf{S}(-\bmq) \cdot \mathbf{S}(\bmq)\rangle $, where $|Q|=1$ is associated with twofold pinch points. Furthermore, expanding $L_m(\bmq)$ in powers of $\bmq$ around $\bmq^\star$ reveals the underlying continuum gauge theory. 

The number of such defects and their arrangement in the Brillouin zone yields a phase diagram spanned by $\alpha$ and $\beta$ featuring 10 distinct classical spin liquids, see \cref{fig:octochlore}(b). In particular, at points along the boundary [i.e., point A in \cref{fig:octochlore}(b)] multiple defects with $Q=\pm1$ merge, leading to a higher $|Q| >1$ associated with a tensor gauge theory and multifold pinch points, see Ref.~\cite{Benton2021}. In addition, we have identified even richer phenomena at crossing points of several phase boundaries: Point B in \cref{fig:octochlore}(b) displays a pinch point with purely parabolic contours, recently predicted to be a hallmark signature of type-II fracton phases~\cite{Hart2022}, while point C features unusual, one-dimensional manifolds of pinch points, so-called pinch-lines \cite{Benton2016}.

{\it Methods---}
The classical model in \cref{eq:OctochloreH} is treated within a standard large-$N$ approach~\cite{Isakov2004}, both at zero and finite temperatures, previously found to correctly encapture the qualitative behaviour for this system \cite{Benton2021}. To study the vastly more complicated quantum $S=1/2$ version, we employ two functional renormalization group (FRG) approaches that replace spin operators by fermionic pseudo particles. An established approach at zero temperature is the so-called pseudo fermion (PF-) FRG \cite{ReutherOrig,Iqbal-2016_3d,Baez2017,Buessen2019,Thoenniss2020,kiese2022,Roscher2018}, in which spin $S=1/2$ operators are mapped onto two flavors of complex fermions $f_{i\uparrow},f_{i\downarrow}$ as $S^\mu_i = \frac{1}{2}\sum_{a,b \in \{\uparrow, \downarrow\}} f^\dagger_{ia} \sigma^\mu_{ab} f_{ib} $. At finite temperatures, we apply the pseudo Majorana (PM-) FRG, where we, instead, represent spins by three flavors $\mu = x,y,z$ of SO(3)-symmetric Majorana fermions $ \{ \eta^\mu_i,\eta^\nu_j \} = \delta_{ij} \delta_{\mu \nu}$ as $ S^\mu_i = -\frac{i}{2} \sum_{\nu,\sigma}\epsilon_{\mu\nu\sigma}\eta^\nu_i \eta^\sigma_i$, without introducing unphysical states \cite{Niggemann2021,Niggemann2022,Schneider2022}.
For both approaches, the resulting interacting model is treated in the thermodynamic limit \footnote{Please see Supplemental Material at end of manuscript for methodological details and a discussion on pinch points.} using one-loop FRG. Here, $\sim 10^8$ first order ordinary differential equations are solved numerically as a function of an artificial Matsubara frequency cutoff $\Lambda$. In the physical limit $\Lambda \rightarrow 0$, we obtain renormalized fermionic vertex functions well beyond mean field, from which we calculate the equal-time spin structure factor $\mathcal{S}(\bm{q})$. Despite the common FRG background, the approximations associated with a one-loop scheme are different in both approaches such that one can consider the PFFRG and PMFRG as independent and complementary techniques. Still, we observe excellent agreement between the equal-time PMFRG structure factor for the lowest simulated temperatures with the one obtained from PFFRG at $T=0$~\cite{Note1}.

{\it Twofold pinch points.---}
Even though not the focus of this work, we start with a brief discussion of more conventional twofold pinch points with $|Q|=1$, occurring in the bulk of every phase of \cref{fig:octochlore}(b). At the pinch point positions $\bmq=\bmq^\star$, the lowest non-vanishing term in an expansion of $L_m(\bmq)$ is the linear one, and hence, the emergent continuum Gauss law has the linear form $\mathbf{\nabla}\cdot\mathbf{E}(\mathbf{r})=\bmq\cdot\mathbf{E}(\bmq)=0$ where $\mathbf{E}(\bmq)=\sum_m S^z_m(\bmq)\partial_{\bmq}L_m(\bmq)$~\footnote{Derivatives here are taken at $\bmq=\bmq^\star$ and $\mathbf{q}$ is meant to be the momentum relative to $\mathbf{q}^\star$. Furthermore, since the different spin components $\mu=x,y,z$ do not couple in the spin constraint, one may restrict to the $z$-component without loss of generality.}. Under the influence of quantum fluctuations at $T=0$ in the $S=1/2$ case treated with PFFRG, twofold pinch points show the typical broadening illustrated in Fig.~\ref{fig:octochlore}(c) for the case $\alpha=\beta=0$, while the overall pinch point shape stays rather intact. In particular, we observe the effects of quantum fluctuations to be analogous to those at a finite temperature $T \sim 1.3$. This broadening indicates violations of the ice rule constraint, and is expected as the absolute spin magnitudes $\mathbf{M}^2_{\textrm{oct},\alpha \beta}$ of neighboring octahedra do not mutually commute and thus fluctuate, i.e., $\langle\mathbf{M}^2_{\textrm{oct},\alpha \beta}\rangle \neq 0$. Importantly, the signal at $\bmq=\bmq^\star$ remains strong and no indications for magnetic long-range order are observed in the full $\alpha$-$\beta$ plane~\cite{Henley2005,Note1}. 
We find these observations to be in direct analogy with past studies of the closely related nearest neighbor pyrochlore Heisenberg model \cite{iqbal19,kiese2022,Niggemann2022,Schafer-2020,Hagymasi-2021,Astrakhantsev-2021,Muller-2019,Derzhko-2020,Hering-2022,Schaefer2022}. 

\begin{figure}
    \centering
    \includegraphics[width = \linewidth]{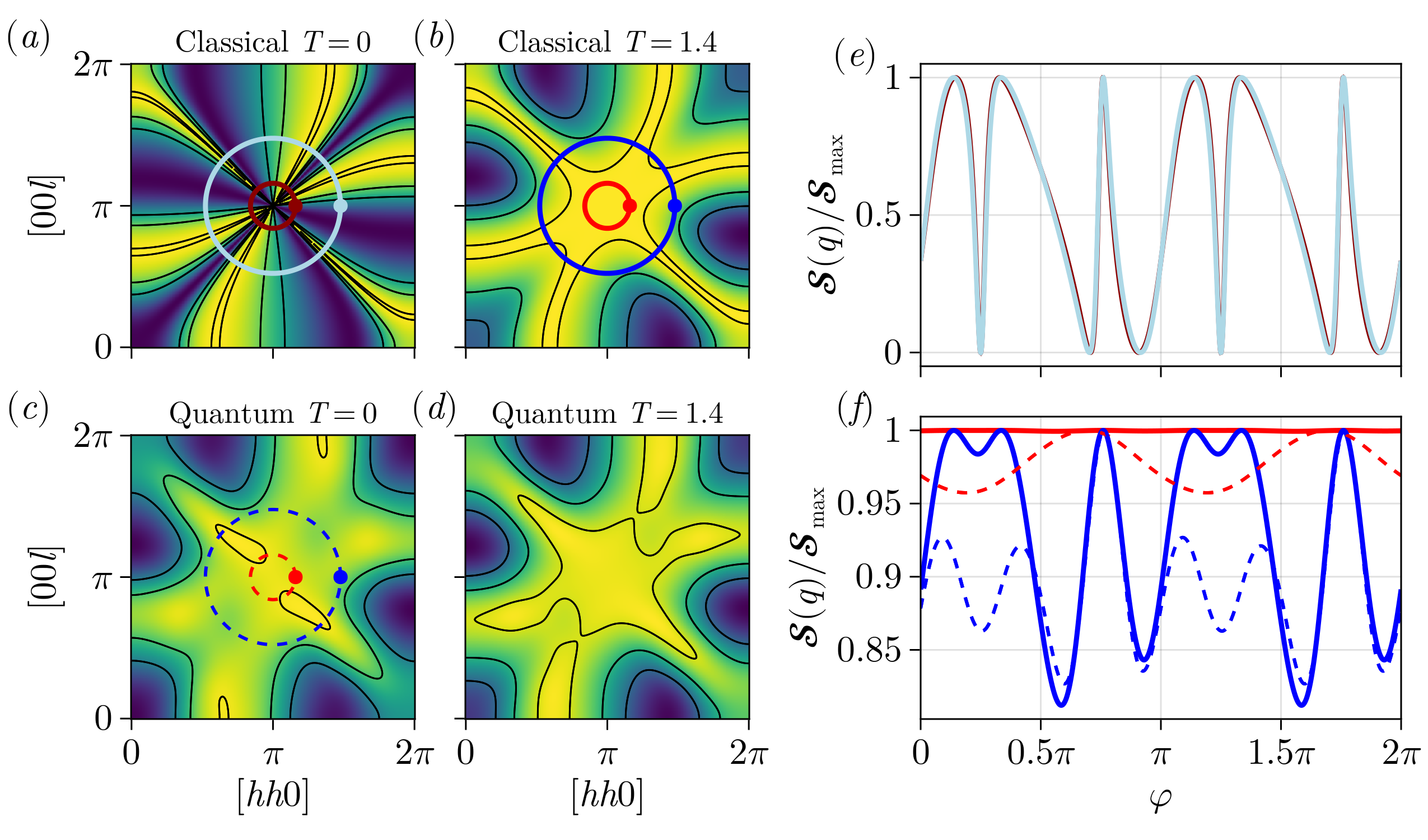}
    \caption{(a-d): Spin structure factor $\mathcal{S}(\bm{q})$ in the $hhl$-plane for a multifold $Q=-7$ pinch point found at location A ($\alpha = -\frac{11}{10}$, $\beta = \frac{9}{5}$) in the phase diagram of \cref{fig:octochlore}. (e): $\mathcal{S}(\bmq)$ for the two paths indicated in (a), normalized to its maximum value along each path. (f): As in (e) but comparing classical thermal and quantum fluctuations along the paths in (b) and (c). The paths are counterclockwise and the start point $\varphi=0$ is indicated by a marker.}
    \label{fig:Multifold}
\end{figure}

{\it Multifold pinch points.---}
A vanishing linear term in an expansion of $L_m(\bmq)$ around $\bmq=\bmq^\star$ is associated with multifold pinch points~\cite{Prem2018,Yan2020,Benton2021}. An instructive example occurs at $\alpha = -11/10$, $\beta = 9/5$ and $\bmq^\star=(\pi,\pi,\pi)$, with a topological index $Q=-7$ and six lobes of large intensity in the $hhl$ plane, see Fig.~\ref{fig:Multifold}(a) and Ref.~\cite{Benton2021}. We identify a gauge constraint of third rank $\sum_{\mu\nu\sigma}q_\mu q_\nu q_\sigma E_{\mu\nu\sigma}(\bmq)=0$ where $E_{\mu\nu\rho}(\bmq) = \sum_m S^z_m(\bmq) \partial_{q_\mu} \partial_{q_\nu} \partial_{q_\rho} L_m(\bmq)$, implying conserved scalar charge, dipole and quadrupole moments. 
\Cref{fig:Multifold} shows the impact of both quantum and thermal fluctuations on this multifold pinch point. 

The value of $\mathcal{S}(\bmq)$ along circular paths around the pinch point illustrates the presence of the singularity: For the exact gauge theory in the classical $T=0$ model, it retains the same strong angular dependence for arbitrarily small radii, see~\cref{fig:Multifold}(e). Thermal fluctuations induce a rather featureless broadening and the angular dependence of the signal vanishes at small distances from the pinch point, see full red line in~\cref{fig:Multifold}(f).
Interestingly, the effects of quantum fluctuations are very different. In addition to a broadening, quantum fluctuations add a shift of spectral weight away from the pinch point origin in favor of soft maxima at incommensurate positions, effectively tearing apart the pinch point. We note that this observation stands in stark contrast to the case of twofold pinch points shown in \cref{fig:octochlore}(c), for which quantum fluctuations appear to act similarly to thermal ones.

\begin{figure}
    \centering
    \includegraphics[width = \linewidth]{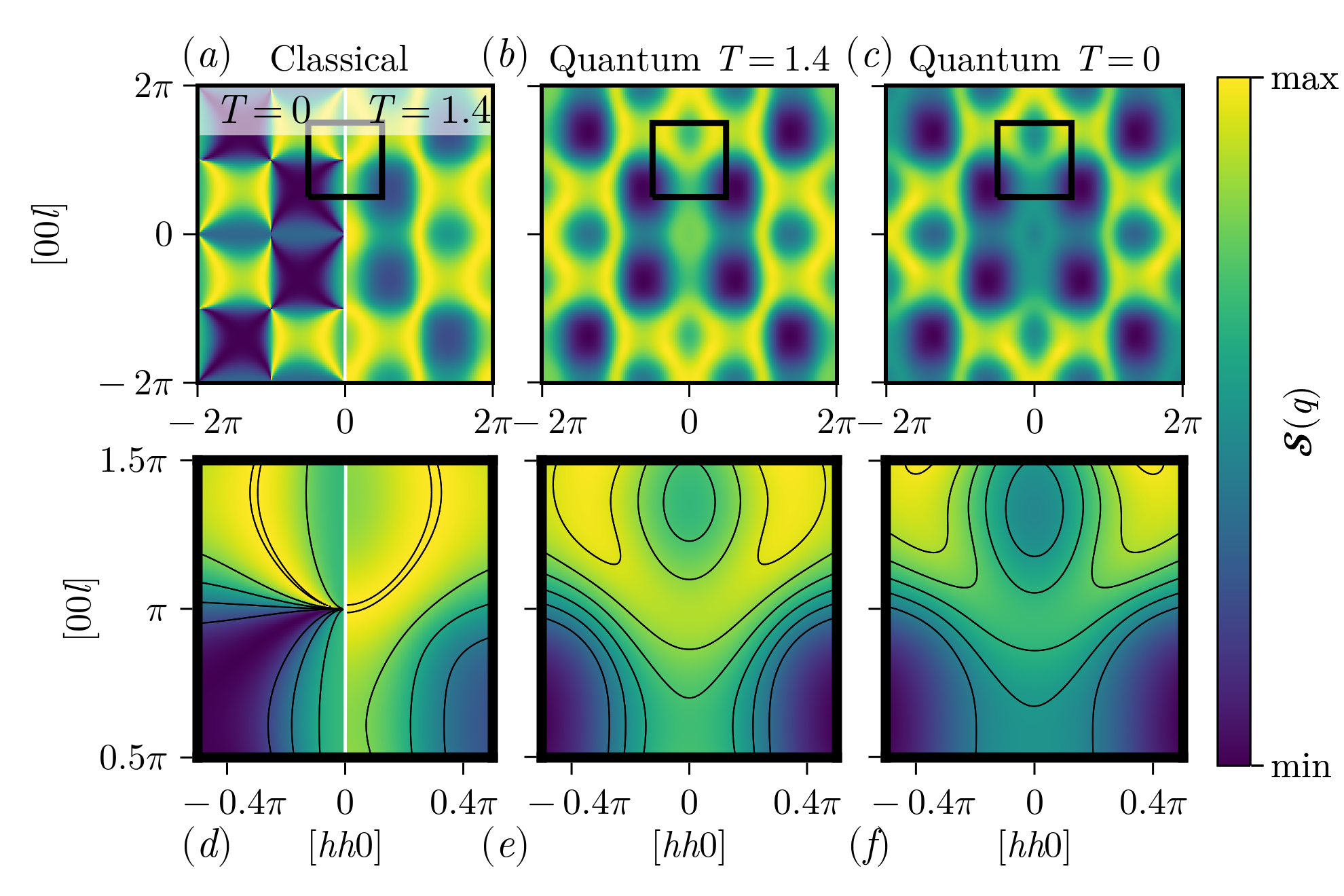}
    \caption{Pinch point with parabolic contours in the $hhl$-plane found at point B ($\alpha = 0, \beta = -1$).
    (a): Classical large-$N$ result for temperatures $T=0$ (left half) and $T=1.4$ (right half). (b) Quantum model at $T=1.4$ and (c) at $T=0$ obtained from PMFRG and PFFRG in the low cutoff limit, respectively. (d-f): Magnifications of the regions indicated by black squares in the upper panel together with black contour lines.}
    \label{fig:Parabolic}
\end{figure}

In order to physically interpret the data in \cref{fig:Multifold}, two types of quantum effects need to be distinguished. First, the aforementioned broadening of pinch points rather indicates the destruction of the underlying gauge theory. However, a second well-known quantum effect consistent with a gauge theory is the formation of gapless photon modes with dispersion $\omega({\bm q})$, resulting from an emergent conjugate vector potential $A({\bm r})$. These photon modes give rise to an extra factor $\omega({\bm q})$ in the spin structure factor (i.e., $\mathcal{S}({\bm q})\rightarrow\omega({\bm q})\mathcal{S}({\bm q})$) suppressing the signal at the singularity due to $\omega({\bm q}^\star)=0$~\cite{benton2012,Prem2018,Hart2022}. To test whether the weight distribution in \cref{fig:Multifold}(c) contains possible signatures of such a modulation, we note that the mere multiplication of an exact pinch point with an isotropic factor $\omega({\bm q})\sim |{\bm q}-{\bm q}^\star|^\gamma$ (or, for that matter, \textbf{any} function $\omega(|{\bm q}-{\bm q}^\star|)$)~\footnote{In the most common types of U(1) gauge theories including rank-1 and rank-2 versions the photon dispersion is linear, i.e., $\gamma=1$, as long as charges are scalar~\cite{Prem2018}.} leaves the singularity intact such that $\mathcal{S}({\bm q})$ along rings around the pinch point, normalized to its maximum on each path, would remain unchanged upon decreasing the radius of the rings. However, the dashed red and blue graphs in \cref{fig:Multifold}(f) illustrating the normalized signal along the ring-like paths in \cref{fig:Multifold}(c) are qualitatively very different and, hence, our results seem incompatible with an emergent photon mode. While it is possible that the ground state is described by a different gauge theory (i.e., with an emergent electric field given by a more complex function of spin operators), we deem it questionable whether fractonic phenomena that have been associated with these spectroscopic features still occur in the $S=1/2$ limit of the Heisenberg model.

{\it Quadratic pinch points.---}
A further generalization occurs if the gauge constraint contains derivatives of different orders as is characteristic for multipolar gauge theories describing type-II fracton phases. This gives rise to quadratic pinch points in the spin structure factor where lobes of strong intensity follow contour lines of the form $q_{\parallel}\propto a q_{\perp}^2$ with $q_{\parallel}$ and $q_{\perp}$ being two perpendicular momentum space directions and $a$ is the lattice constant (which is set to one here). The mixing of derivatives causes the lattice constant to explicitly appear in these spectroscopic patterns which is a direct manifestation of the ultraviolet-infrared mixing described in recent literature~\cite{Hart2022}.

Strikingly, we have identified such quadratic pinch points in the classical octochlore model at $\alpha=0$, $\beta=-1$ and $\bmq^\star = (0,0,\pi)$. The effective gauge theory in this case contains first derivatives along the $z$-direction, as $\partial_{q_z}L_3(\bmq)\neq0$, while for the perpendicular $x$, $y$ directions $\partial_{q_x} L_m(\bmq) = \partial_{q_y} L_m(\bmq) =0$ for $m=1,2,3$ and the lowest non-vanishing contribution comes from second derivatives. The resulting quadratic pinch point in classical large-$N$ [\cref{fig:Parabolic}(d)] has a shape which is similar to predictions from the U(1) Haah code~\cite{Hart2022}. The effect of finite temperatures in large-$N$ only amounts to a broadening near $\bmq^\star$ while retaining the quadratic shape and the strong signal around $\bmq^\star$. This is to be contrasted with PMFRG at the same temperature where the signal is reduced near $\bmq^\star$ and quadratic contours are no longer discernible. This trend continues down to $T=0$ where the spin-structure factor appears even more strongly reduced around $\bmq^\star$. Again however, this result seems incompatible with emergent photons, see Ref.~\cite{Note1}.

As a side remark, the model with $\alpha=0$, $\beta=-1$ also hosts a fourfold pinch point~\cite{Yan2020}, see Fig.~\ref{fig:Parabolic}(a) at $\bmq^\star=\mathbf{0}$, associated with a trace-full rank-2 tensor gauge constraint. With the observed reduction of $\mathcal{S}(\mathbf{q})$ and the formation of a local minimum at $\bmq^\star=\mathbf{0}$ in the quantum model as $T\rightarrow0$, this is another example reflecting the strong impact of quantum fluctuations on exotic pinch points. 

\begin{figure}
    \centering
    \includegraphics[width = \linewidth]{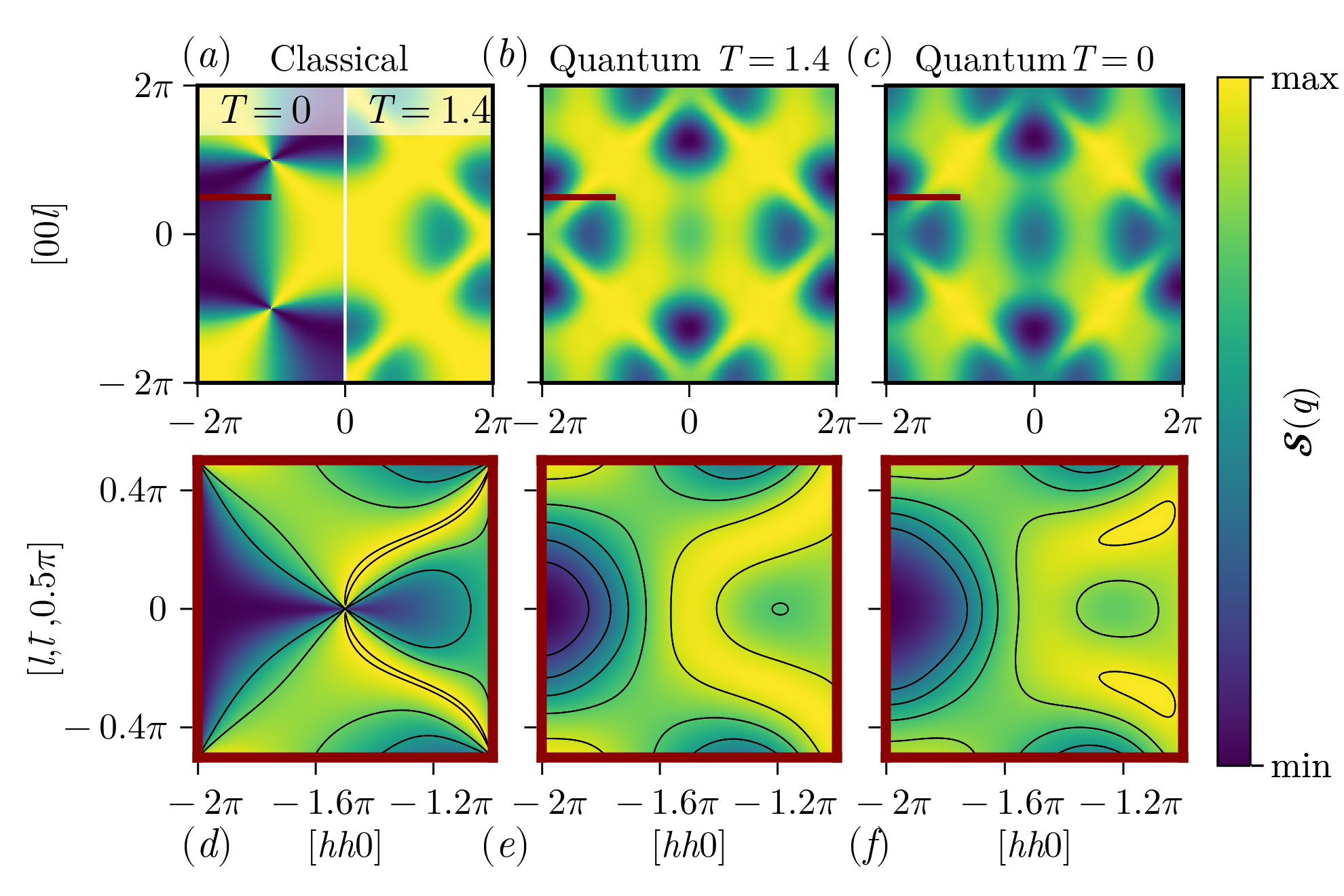}
    \caption{(a-c): Temperature dependent spin structure factor for a pinch line at $\alpha = -\frac{1}{2}$, $\beta = 1$
    [C in \cref{fig:octochlore}(b)] in the $[111]$ direction. Panels (d-f) show a cut through the pinch line, here given by $q_z = 0.5\pi$ as indicated by the solid dark red line in panels (a-c).}
    \label{fig:Pinchline}
\end{figure}

{\it Pinch-lines.---}
Points $\bmq^\star$ of vanishing constraint vector are not necessarily isolated in momentum space but can form one-dimensional manifolds. This situation has previously been studied in Ref.~\cite{Benton2016} where the phenomenon has been dubbed a pinch line. Such patterns exhibit conventional twofold pinch points in all planar cuts through the pinch line. For the classical pyrochlore model investigated in Ref.~\cite{Benton2016} an underlying gauge constraint linear in the derivatives but with a tensor structure has been identified and a possible relevance for the pyrochlore material \ce{Tb2Ti2O7} \cite{Fennell2012} has been pointed out.

We have found an analogous feature in the octochlore model at $\alpha=-\frac{1}{2}$ and $\beta=1$ where pinch lines run along $[111]$ and symmetry related directions in momentum space. The lowest non-vanishing derivatives of $L_m(\bmq)$ at $\bmq^\star$ are first order derivatives perpendicular to the pinch lines, in agreement with the pyrochlore model of Ref.~\cite{Benton2016}. Since the topological defect is now line-like and observing that the normalized constraint vector $\tilde{L}_m(\bmq)$ avoids two opposite points on the unit sphere $S^2$~\cite{Note1}, the topological index is given by the integer vortex winding number $w$. We find $|w|=1$ and consequently, twofold pinch points in planar cuts through the line defect, see bottom panel of Fig.~\ref{fig:Pinchline} depicting cuts at $q_z=0.5\pi$. Thermal fluctuations in the classical model [Fig.~\ref{fig:Pinchline}(a), right] shift spectral weight towards the pinch lines such that they become visible in the $hhl$ plane as well defined, broadened lines of constant strong signal.

For the corresponding quantum model, similar observations to the previous cases can be made, such as a re-distribution of spectral weight away from the pinch line when temperature is lowered, as shown in the bottom panel of Fig.~\ref{fig:Pinchline}. It is again worth contrasting this behavior with conventional twofold pinch points representing isolated point defects where quantum fluctuations are not seen to significantly reduce the signal at $\bmq=\bmq^\star$, see the example in Fig.~\ref{fig:octochlore}(c).

{\it Discussion.---}
We have identified the classical octochlore model as an exquisite physical platform for studying exotic spectroscopic features, such as multifold and quadratic pinch points as well as pinch lines, all associated with unconventional gauge theories. Numerical studies which systematically investigate the impact of quantum fluctuations on the corresponding classical spin liquids are, however, lacking so far. In our endeavor to fill this gap, we treat the quantum spin $S=1/2$ model employing state-of-the-art PFFRG and PMFRG methods. We find a recurring theme in our results: Multifold pinch points, quadratic pinch points and pinch lines all undergo a significantly different modification under quantum fluctuations than conventional twofold pinch points, showing a reduction of $\mathcal{S}(\mathbf{q})$ at $\mathbf{q}^\star$ that is also at variance compared to the effects of pure thermal fluctuations in the classical case. This also implies that the absence of unconventional pinch points in an experimentally measured spin structure factor does not necessarily exclude the realization of a higher-rank $U(1)$ gauge theory in the corresponding classical system.

From a methodological perspective, here, we benefit from the fact that our octochlore model has SU(2) spin symmetry which simplifies the application of PFFRG and PMFRG enormously. A (numerically more challenging) continuation of our present work could be to lift the SU(2) symmetry by considering an Ising version of the octochlore model supplemented with small transverse couplings, thus realizing an analogous situation as in quantum spin ice models. This will help identifying the fate of exotic pinch point singularities along a continuous classical-to-quantum interpolation. 
Thus, our results strongly motivate new avenues in the investigations of these exotic pinch points under quantum fluctuations, which appear to have a more significant impact compared to twofold-pinch points. Furthermore, our work sets the stage for determining the microscopic wave functions describing these resulting quantum phases, and whose correlation functions give rise to the static structure factors obtained here~\cite{Hering-2019}.

{\it Acknowledgements.---}
We thank Owen Benton and Roderich Moessner for insightful discussions.
N.\,N. and J.\,R. acknowledge support from the Deutsche Forschungsgemeinschaft (DFG, German Research Foundation), within Project-ID 277101999 CRC 183 (Project A04).
N.\,N. thanks IIT Madras for funding a three-month stay through an International Graduate Student Travel award which facilitated completion of this research work.
J.\,R. thanks IIT Madras for a Visiting Faculty Fellow position under the IoE program during which part of the research work and manuscript writing were carried out.
Y.~I.~acknowledges support from DST, India through MATRICS Grant No.~MTR/2019/001042, CEFIPRA Project No.~64T3-1, ICTP through the Associates Programme and from the Simons Foundation through grant number 284558FY19. This research was supported in part by the National Science Foundation under Grant No.~NSF~PHY-1748958, IIT Madras through the Institute of Eminence (IoE) program for establishing the QuCenDiEM group (Project No. SP22231244CPMOEXQCDHOC), the International Centre for Theoretical Sciences (ICTS), Bengaluru, India during a visit for participating in the program ``Frustrated Metals and Insulators'' (Code: ICTS/frumi2022/9).
N.~N. acknowledges usage of the JUWELS cluster at the Forschungszentrum J\"ulich and the Noctua2 cluster at the Paderborn Center for Parallel Computing (PC$^2$). Y.~I.~acknowledges the use of the computing resources at HPCE, 
IIT Madras.
\bibliography{higher_rank}

\newcommand{\beginsupplement}{%
        \setcounter{table}{0}
        \renewcommand{\thetable}{S\arabic{table}}%
        \setcounter{figure}{0}
        \renewcommand{\thefigure}{S\arabic{figure}}%
        \setcounter{equation}{0}
        \renewcommand{\theequation}{S\arabic{equation}}%
        \setcounter{page}{1}
     }

\renewcommand*{\citenumfont}[1]{S#1}
\renewcommand*{\bibnumfmt}[1]{[S#1]}
 
\newcommand\blankpage{%
    \null
    \thispagestyle{empty}%
    \addtocounter{page}{-1}%
    \newpage}

\blankpage
\blankpage

\chead{{\large \bf{---Supplemental Material---}}}

\thispagestyle{fancy}

\beginsupplement

\maketitle

\section*{Emergent gauge theories}
The purpose of this suppelementary section is to provide a more in-depth introduction to the construction of emergent gauge theories on the octochlore model.  
As discussed in Ref.~\cite[]{Benton2021}, the classical ground state constraint of the octochlore model can be written as
\begin{align}
    \sum_{i \in c} \eta_i \bm{S}_i &= 0 \quad \forall c,\\
    \eta_i &= \begin{cases}
        1, \quad i \in \textrm{oct}\\
        \alpha, \quad i \in \langle\textrm{oct} \rangle\\
        \beta, \quad i \in \langle\langle\textrm{oct} \rangle\rangle\\
    \end{cases}
    ,
\end{align}
where $c$ is the cluster of octahedra shown in Fig.~1 of the main text.
In reciprocal space, the constraint can be expressed using a constraint vector $L_m(\bmq)$
\begin{align}
    L_m(\bmq) &= \sum_{i \in m \in c} \eta_i  e^{\imath \bmq (\bm{r}_c - \bm{r}_i)}\\
    \sum_{m=1}^{n_u} L^*_m(\bmq) \bm{S}_m(\bmq) &= 0 \quad \forall \bmq,\label{eq:qSpaceConstraint}
\end{align}
where $n_u=3$ is the number of sites per unit cell, $\bm{r}_c$ indicates the position of the center of the cluster $c$ and $\bm{r}_i$ the position of site $i$.
As all spin components are equivalent, henceforth, we only consider the $z$-component.
Even though the dimension of $L_m$ is given by the number of sublattices and can in principle be of arbitrary dimension, here we shall label its three components as $L_x,L_y,L_z$ for notational convenience.
\Cref{eq:qSpaceConstraint} implies that the vector $S^z_m$ is orthogonal to the constraint vector $L_m(\bmq)$. Hence, the spin structure factor $\mathcal{S}(\bmq) \equiv \frac{1}{n_u} \sum_{m,n} \langle S^z_m(-\bmq) S^z_n(\bmq) \rangle$ can be approximated at zero temperature by summing over all elements of the matrix projecting out $\tilde{L}_m(\bmq)$ \cite[]{Henley2005,Benton2021}. This explains the appearance of pinch points whenever $L_m(\bm{q}) = \bm{0}$ and the projector becomes singular.
The effective gauge theory is then given by expanding $L_m(\bmq)$ to leading order around the location of a pinch point $\bmq^\star$, corresponding to a coarse graining of the system. If the lowest non-vanishing contribution is of first order, we obtain
\begin{align}
  \sum_m \sum_{\mu} \left. \frac{\partial L^*_m}{\partial \tilde{q}_\mu} \right|_{\tilde{\bmq} = 0} \tilde{q}_\mu S^z_m(\tilde{\bmq}) \equiv \sum_{\mu} \tilde{q}_\mu  E_\mu(\tilde{\bmq}) = 0,
\end{align}
where $\tilde{\bmq} = \bmq - \bmq^\star$.
This is a simple Gauss' law $\nabla \cdot \bm{E} = 0$ in reciprocal space. The emergent gauge field $E_\mu(\tilde{\bmq}) = \sum_m \frac{\partial L^*_m}{\partial \tilde{q}_\mu}|_{\tilde{\bmq}= 0} S^z_m(\tilde{\bmq})$ in this example is of rank-1 U(1) type. An interesting special case emerges when the gradient of the constraint vector also vanishes. In this case, the effective gauge field becomes a higher rank tensor which may depend on terms such as $ \frac{\partial^2 L^*_m}{\partial q_\mu \partial q_\nu}$.
We now consider more explicit examples found on the octochlore model, for which the constraint vector can be written as
\begin{widetext}
    \begin{equation}
        \bm{L}(\bmq) = 
        \begin{pmatrix} 2 \cos \left(\frac{q_x}{2}\right) \left[ 2 \alpha  (\cos (q_y)+\cos (q_z))+2 \beta  \cos (q_x)+1-\beta \right]\\
            2 \cos \left(\frac{q_y}{2}\right) \left[ 2 \alpha  (\cos (q_x)+\cos (q_z))+2 \beta  \cos (q_y)+1-\beta \right]\\
            2 \cos \left(\frac{q_z}{2}\right) \left[ 2 \alpha  (\cos (q_x)+\cos (q_y))+2 \beta  \cos (q_z)+1-\beta \right]
        \end{pmatrix}\label{eq:LOctochlore}.
    \end{equation}
\end{widetext}

{\it Twofold pinch point.---} 
First, consider the simple special case $\alpha = \beta =0$. In this case we have $\bm{L} = 2 \left(\cos(q_x/2),\cos(q_y/2),\cos(q_z/2)\right)$ which vanishes only at the pinch point $q^\star = (\pi,\pi,\pi)$ (and equivalent positions). Here, we find $\partial_{q_\mu} L_m(\bmq^\star) = -\delta_{\mu m}$, i.e., the underlying gauge structure can be described by an emergent rank-1 gauge field, as expected.

\begin{figure}
    \centering
    \includegraphics[width = \linewidth]{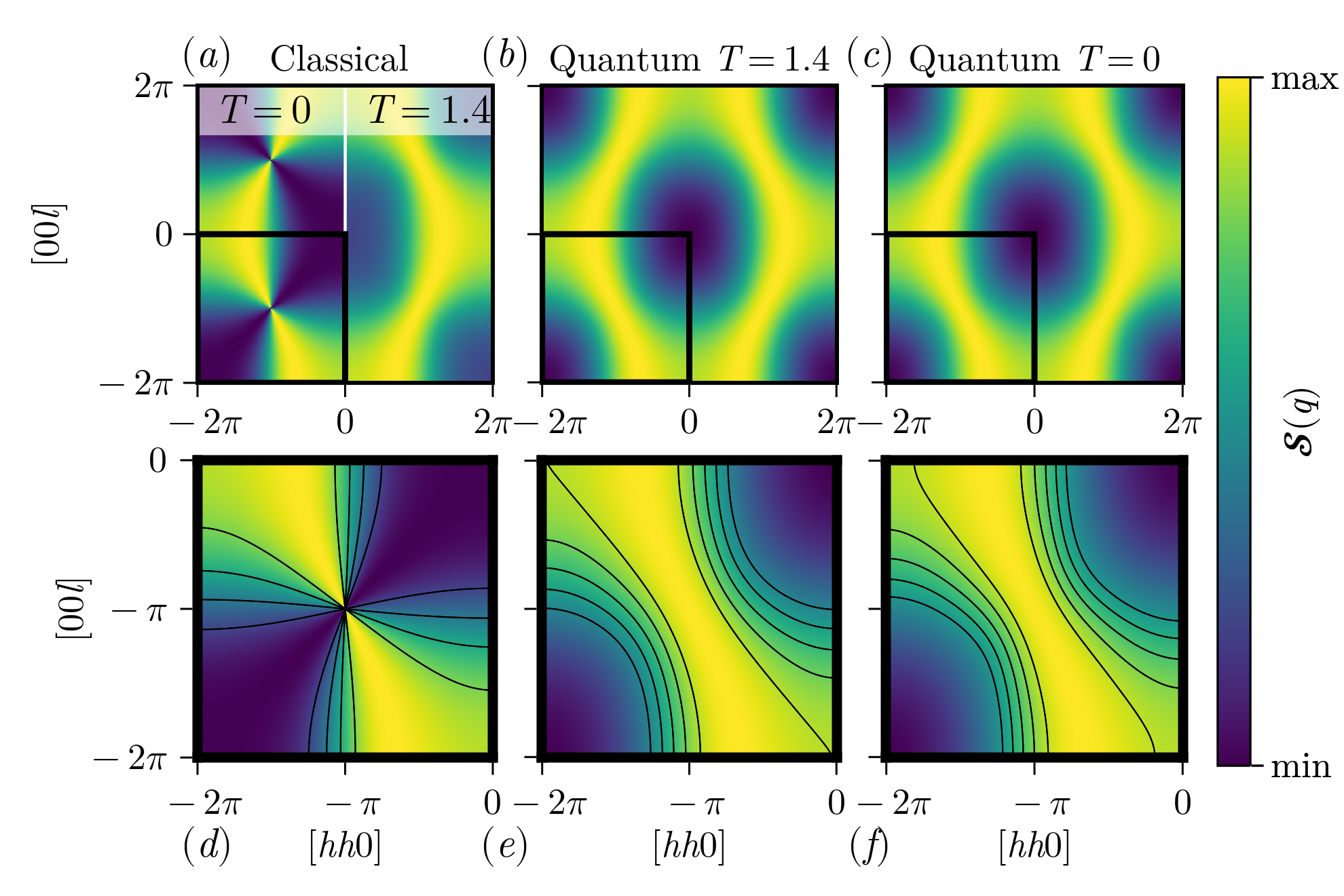}
    \caption{\textbf{Conventional twofold pinch point} at $\alpha = \beta =0$. The arrangement and description of the individual panels corresponds to that in Fig.~3 of the main text.}
    \label{fig:Twofold}
\end{figure}

As argued in the main text, the impact of quantum fluctuations on the unconventional pinch point features differs significantly from conventional ones, such as in the pyrochlore lattice. Here, we demonstrate that we find the same for the octochlore lattice by choosing $\alpha = \beta = 0$. Figs.~\cref{fig:Twofold} and Fig.~1(c,d) of the main text show that for the octochlore model quantum fluctuations only lead to a broadening of the pinch points, while crucially, no relative reduction of spectral weight at the original pinch point location is observed. In particular, it can be seen that the peak at the former pinch point becomes narrower as temperature is lowered, in agreement with classical expectations, analogous to previous findings on the pyrochlore lattice \cite{Niggemann2022}. As argued in the main text, this observation of broadening is analogous to the effect of thermal fluctuations

{\it Fourfold pinch point.---} 
For $\alpha = 0, \beta = -1$, $\bm{L}(\bmq)$ simplifies to 
\begin{equation}
    \bm{L}(\bmq) = 
    \begin{pmatrix}
        4 \cos \left(\frac{q_x}{2}\right)(1-\cos(q_x)) \\
        4 \cos \left(\frac{q_y}{2}\right)(1-\cos(q_y)) \\
        4 \cos \left(\frac{q_z}{2}\right)(1-\cos(q_z)) 
    \end{pmatrix}.
\end{equation}
We find that all components of $\bm{L}(\bmq)$ and its first derivatives vanish at $\bmq^\star = (0,0,0)$. The first nonzero contributions are all of second order
\begin{equation}
    \left.\partial _{q_\mu}\partial _{q_\nu}L_m\right|_{\bmq = \bmq^\star} = 4\delta_{\mu\nu} \delta_{\mu m}
\end{equation}

As the underlying gauge field is a rank-2 tensor $E_{\mu\nu}(\bmq) = 4 \delta_{\mu \nu}  S^z_\mu(\bmq) $, the structure factor displays a fourfold pinch point.
A gauge theory of this form $\partial_\mu \partial_\nu E_{\mu\nu}=0$ implies the existence of quasiparticle excitations with conserved dipole moment \cite[]{Pretko2017}. These so-called \emph{fractons} are thus immobile unless grouped together to form pairs or larger clusters.

\begin{figure}
    \centering
    \includegraphics[width = \linewidth]{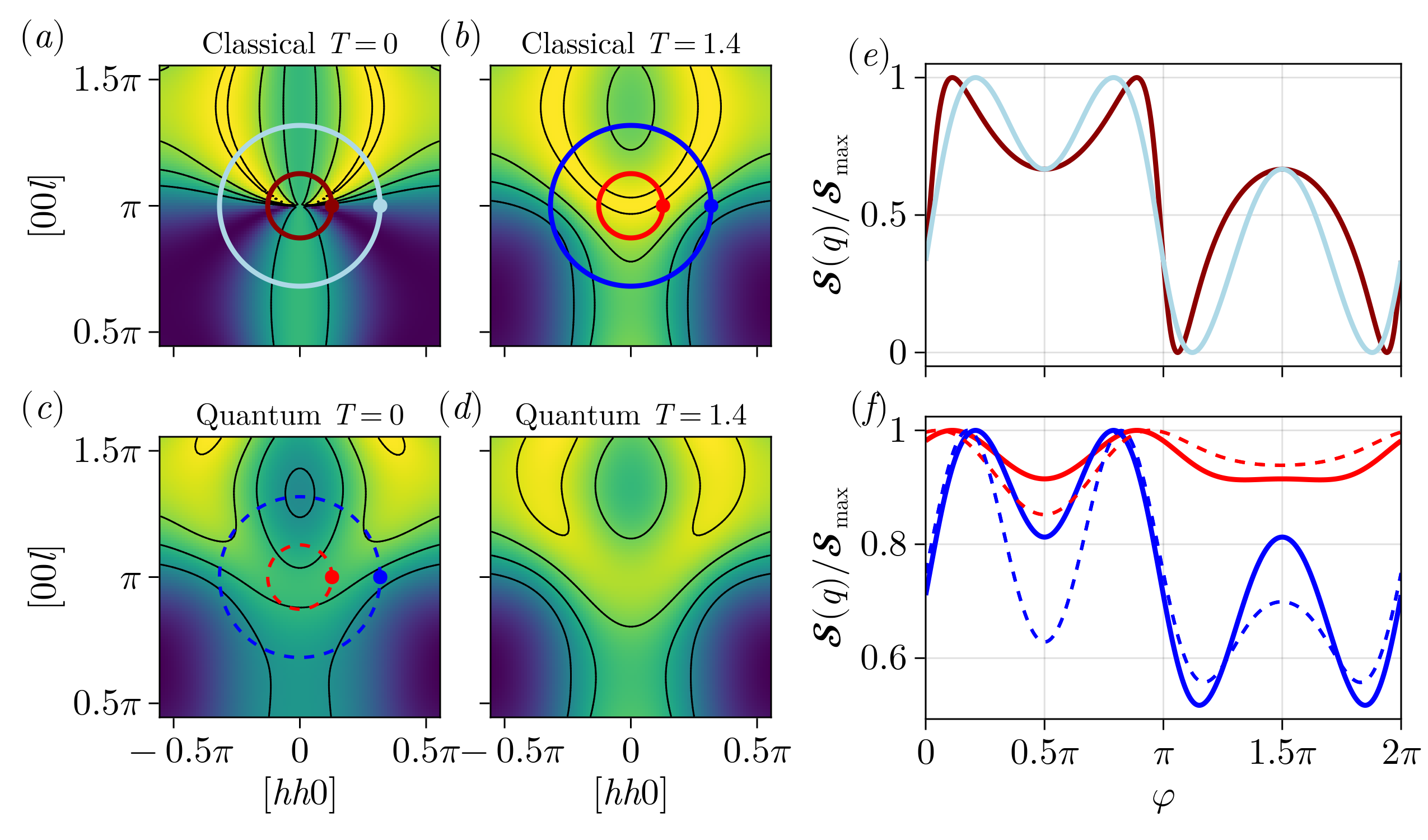}
    \caption{\textbf{Difference between thermal and quantum fluctuations for quadratic pinch point} at $\alpha = 0, \beta = -1$.}
    \label{fig:ComparisonQuadPP}
\end{figure}

{\it Quadratic pinch point.---} 
At the same location in the phase diagram ($\alpha = 0, \beta = -1$), we also find a pinch point with purely parabolic contours at $\bmq^\star = (0,0,\pi)$.
As pointed out by Hart {\it et. al.} in Ref.~\cite[]{Hart2022}, such a pinch point arises from the presence of mixed derivatives in the Gauss law constraint and is a signature of a type-II fractonic phase. Indeed, we verify that for the present case,
$\bm{L}$ has a nonzero first derivative only in the $q_z$ direction: 
\begin{align}
    \left.\partial_{q_\mu} L_m(\bmq)\right|_{\bmq = \bmq^\star} &= \begin{cases}
        -4\quad \mu=m = z\\
        0\quad \text{else}
    \end{cases}\\
    \left.\partial_{q_\mu}\partial_{q_\nu} L_m(\bmq)\right|_{\bmq = \bmq^\star} &= \begin{cases}
        4\quad \mu=\nu=m \text{ and } m=x,y\\
        0\quad \text{else}
    \end{cases}.
\end{align}
\Cref{fig:ComparisonQuadPP} shows a comparison between thermal and quantum fluctuations on a quadratic pinch point. In contrast to the twofold pinch point shown in Fig.~1(c,d) of the main text, the effects of quantum and thermal fluctuations are quite distinct, leading to a suppression of the structure factor around the pinch point.
As discussed in the main text, for systems with emergent photon excitations, a similar suppression is also observed \cite{benton2012}. Here, the structure factor simply acquires a prefactor from the dispersion of a photon with the speed of light $c$. $\omega(\tilde{\bmq}) = c \sqrt{\tilde{q}^2_z + \left(\tilde{q}_x^2+\tilde{q}_y^2\right)^2}$, where  \cite{benton2012}
\begin{equation}
    \mathcal{S}(\tilde{\bmq}) \rightarrow \omega(\tilde{\bmq}) \coth \left( \frac{\omega(\tilde{\bmq})}{2T} \right) \mathcal{S}(\tilde{\bmq}) . \label{eq:photonStructureFactor}
\end{equation} 

\begin{figure}
    \centering
    \includegraphics[width = \linewidth]{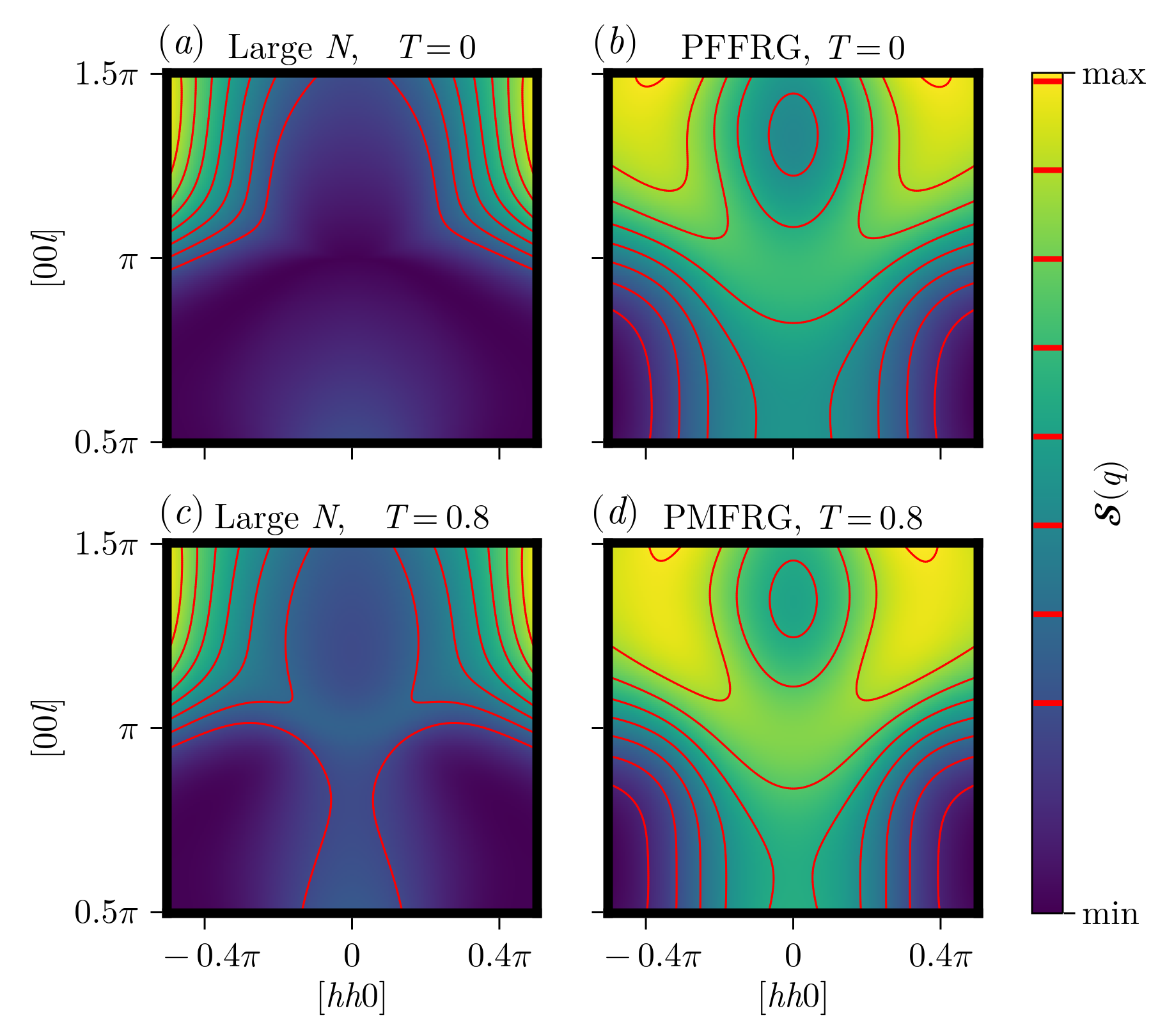}
    \caption{\textbf{Quadratic pinch point at $T=0$ and $T=0.8$}. Dispersion-corrected structure factor $\mathcal{S}(\bmq)$ obtained from large $N$ and \cref{eq:photonStructureFactor} in (a) and (c) in comparison of to PFFRG (b) and PMFRG (d).}
    \label{fig:Quad_Dispersioncorrected}
\end{figure}

In \cref{fig:Quad_Dispersioncorrected}, the effect of such a modification is considered in comparison to the findings from PFFRG. The resulting structure factors are clearly qualitatively distinct. While PFFRG is formally employed at zero temperature, the influence of its finite cutoff is often similar to a finite temperature. To investigate this possibility, a rough estimate of the structure factor can be obtained from large $N$ at finite temperature by applying the correction from the photon dispersion in \cref{eq:photonStructureFactor}. Assuming that $c = 1$ (in units of the normalized lattice constant and energy scale), the result is shown in panel (c) of \cref*{fig:Quad_Dispersioncorrected}. As in the case of quantum spin ice, this returns some spectral weight back to the pinch point. Although the intensity at the pinch point depends on microscopic details such as the value of $c$, we notice that the position of the maxima remains indifferent, following the original parabolic contour. This contrasts our numerical observations so that the presence of emergent photons remains unlikely. 

\begin{figure}
    \centering
    \includegraphics[width = \linewidth]{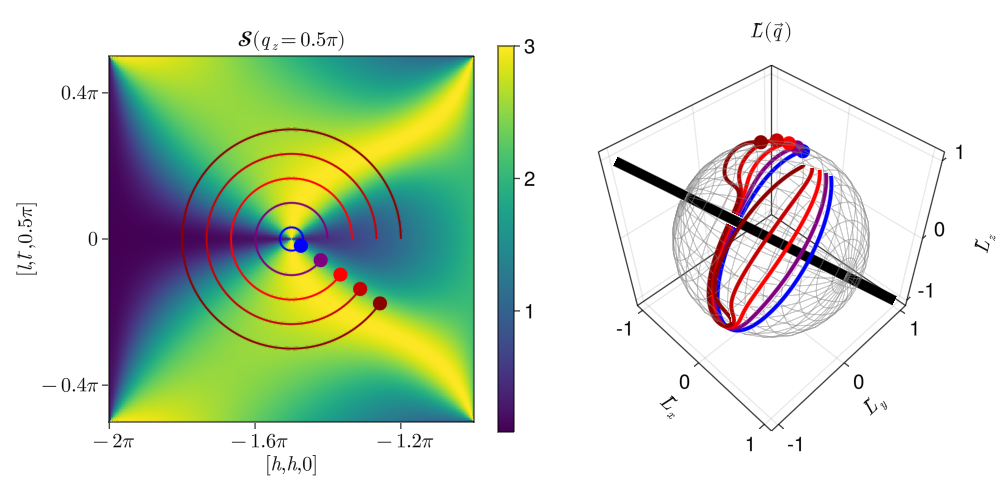}
    \caption{\textbf{Vortex winding number for pinch line.} Left: Spin structure factor from large-$N$ at $T=0$ in a cut through the pinch line, with loops of different radius. Right: For each loop, the normalized constraint vector describes a path winding once around the same axis. Sufficiently close to the pinch point, this path is of circular shape.}
    \label{fig:WindingNumber}
\end{figure}

{\it Multifold pinch point.---} 
At $\alpha = -11/10, \beta = 9/5$, an even higher rank gauge theory of the form $\partial_\mu \partial_\nu \partial_\rho E_{\mu\nu\rho} = 0$ emerges, since both the first and the second order derivatives of the constraint vector vanish at the pinch point $\bmq^\star = (\pi,\pi,\pi)$. Explicitly, we find for the first component of the constraint vector
\begin{align}
    \partial_{q_x}^3 L_x(\bmq^\star)  &= -54/5,\nonumber\\  \partial_{q_x}\partial_{q_y}^2 L_x(\bmq^\star)  &= \partial_{q_x}\partial_{q_z}^2 L_x(\bmq^\star) = 11/5,
\end{align}
while other derivatives such as $\partial_{q_x}^2\partial_{q_y} L_x(\bmq^\star)$ are zero. Consequently, not only monopole and dipole moments, but also the quadrupole moment are conserved. This feature is characterized by a skyrmion winding number of $Q=-7$ and displays a sixfold pinch point when cut through the $hhl$ plane.

{\it Pinch line.---} 
As mentioned before, singularities in the structure factor are present at points $\bmq^\star$ where $L_m(\bmq^\star) = 0$. Inspecting \cref{eq:LOctochlore}, we immediately observe that a pinch point can always be found at $(\pi,\pi,\pi)$, where the prefactors $\cos(q_\mu/2)$ vanish. Other pinch points emerge at more complicated positions determined by a delicate balance between the parameters $\alpha,\beta$ and the wavevector $\bmq$.
In the generic case, for a fixed set of $\alpha, \beta$, the requirement that all three components of $\bm{L}$ have to vanish leads to three equations, determining the positions of the pinch points $\bmq^\star$ uniquely (up to point group symmetries). However, for appropriate $\alpha$ and $\beta$, one or more components can become equivalent, leading to a line-like manifold of singularities.

\begin{figure}
    \centering
    \includegraphics[width = \linewidth]{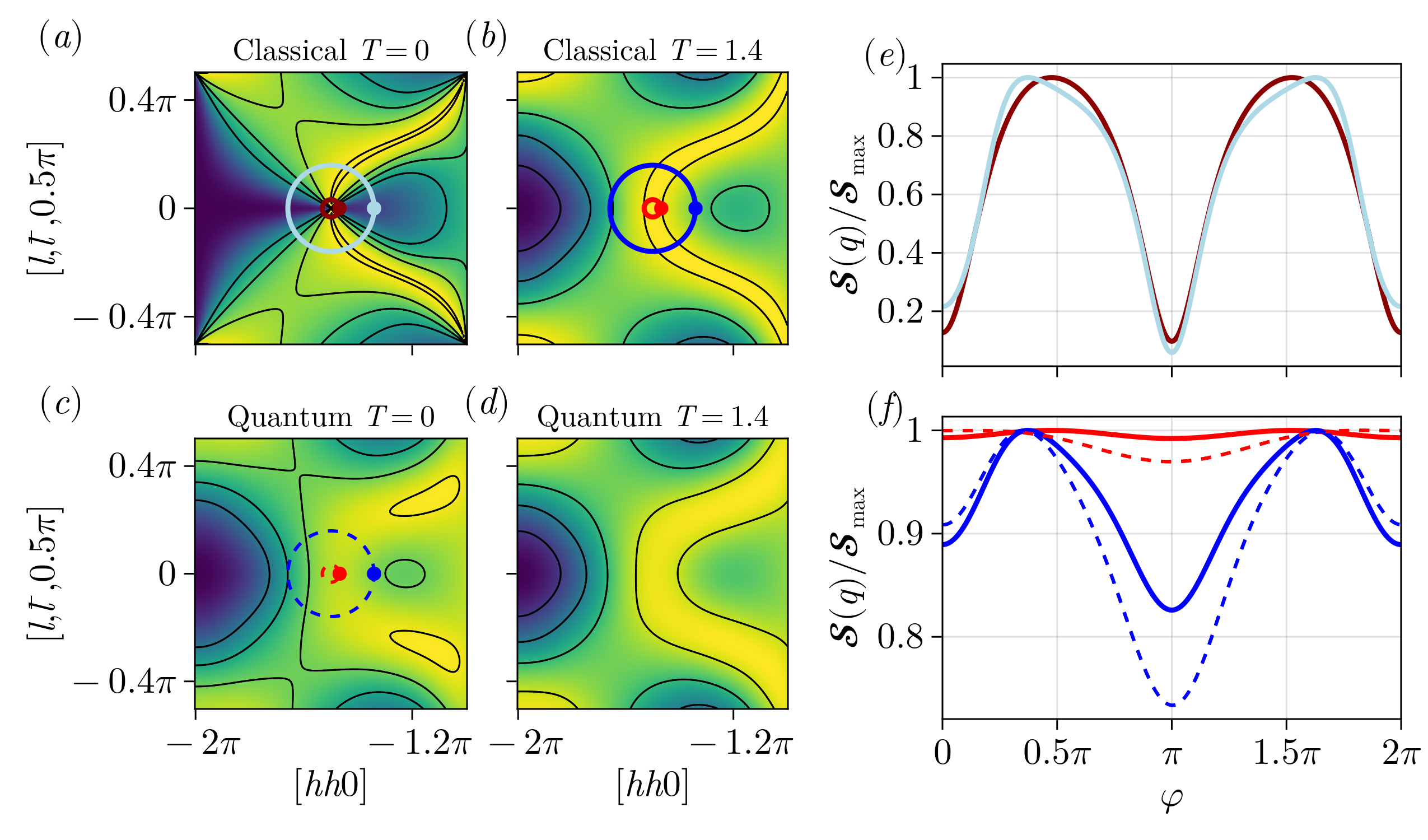}
    \caption{\textbf{Difference between thermal and quantum fluctuations for pinch-line} at $\alpha = -\beta = 1$  .}
    \label{fig:ComparisonPinchLine_circle}
\end{figure}

A simple example is found by setting $q_z=\pi$, such that $L_z = 0$. One can see that whenever $q_x=q_y$, the first two components of $L_m$ become equivalent:
\begin{equation}
    \bm{L} = \begin{pmatrix}
        2 \cos \left(\frac{q_x}{2}\right) \left[\cos \left(q_x\right)\left(2\alpha+2 \beta \right)+1 -2\alpha -\beta \right]\\
        2 \cos \left(\frac{q_x}{2}\right) \left[\cos \left(q_x\right)\left(2\alpha+2 \beta \right)+1 -2\alpha -\beta \right]\\
        0
    \end{pmatrix}
\end{equation}
These two components vanish for all $q_x=q_y$ for $\alpha = -\beta = 1$, resulting in a line of pinch points, or pinch-line. This pinch line can be characterized by a topological, winding number. Here, $\tilde{L}_m(\bmq) = L_m(\bmq) / \sqrt{\sum_n{L_n^2(\bmq)}}$ is traced on the unit sphere as one moves along a closed loop around the pinch line. Since the corresponding paths on the unit sphere avoid two opposite poles (see \cref{fig:WindingNumber}), the topological index can be defined as the corresponding winding number.
For each point on the pinch line, the winding number takes the same integer value as long as the loop does not contain or intersect any other pinch point (where $\tilde{L}(\bmq)$ is singular). \Cref{fig:ComparisonPinchLine_circle} shows that the effects of quantum and thermal fluctuations for one of the pinch points within the manifold. Being of twofold nature, the effects of quantum fluctuations are again, relatively similar to thermal ones, although a shift of spectral weight away from the pinch point center is again observed.

\begin{figure}
    \centering
    \includegraphics[width = 0.5\linewidth]{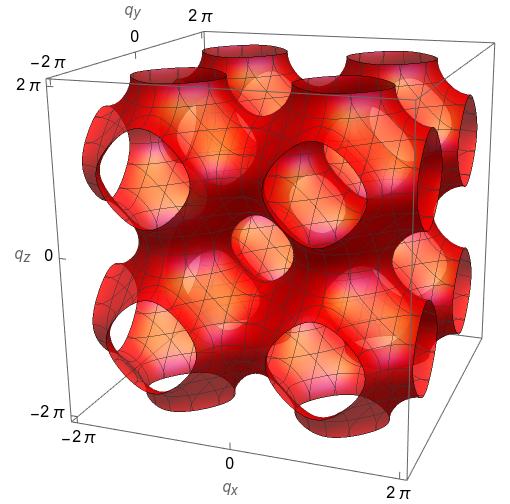}
    \caption{\textbf{Surface with vanishing constraint vector for $\alpha=\beta=2$.}}
    \label{fig:LzeroSurface}
\end{figure}

Finally, we note that it is also possible for all three components of $L_m(\bmq)$ to become equivalent, leading to a surface in reciprocal space for which $L_m(\bmq)=0$. One such example is found at $\alpha = \beta = 2$. The resulting surface of vanishing $L_m(\bmq)$ is displayed in \cref{fig:LzeroSurface}. Although this surface can be found spectroscopically as a narrow local maximum of the structure factor at finite temperature or under the inclusion of quantum fluctuations, becoming infinitely thin as fluctuations decrease, no pinch points are visible since any cut always contains a one-dimensional sub-manifold of this surface.

\begin{figure}
    \centering
    \includegraphics[width = 0.9\linewidth]{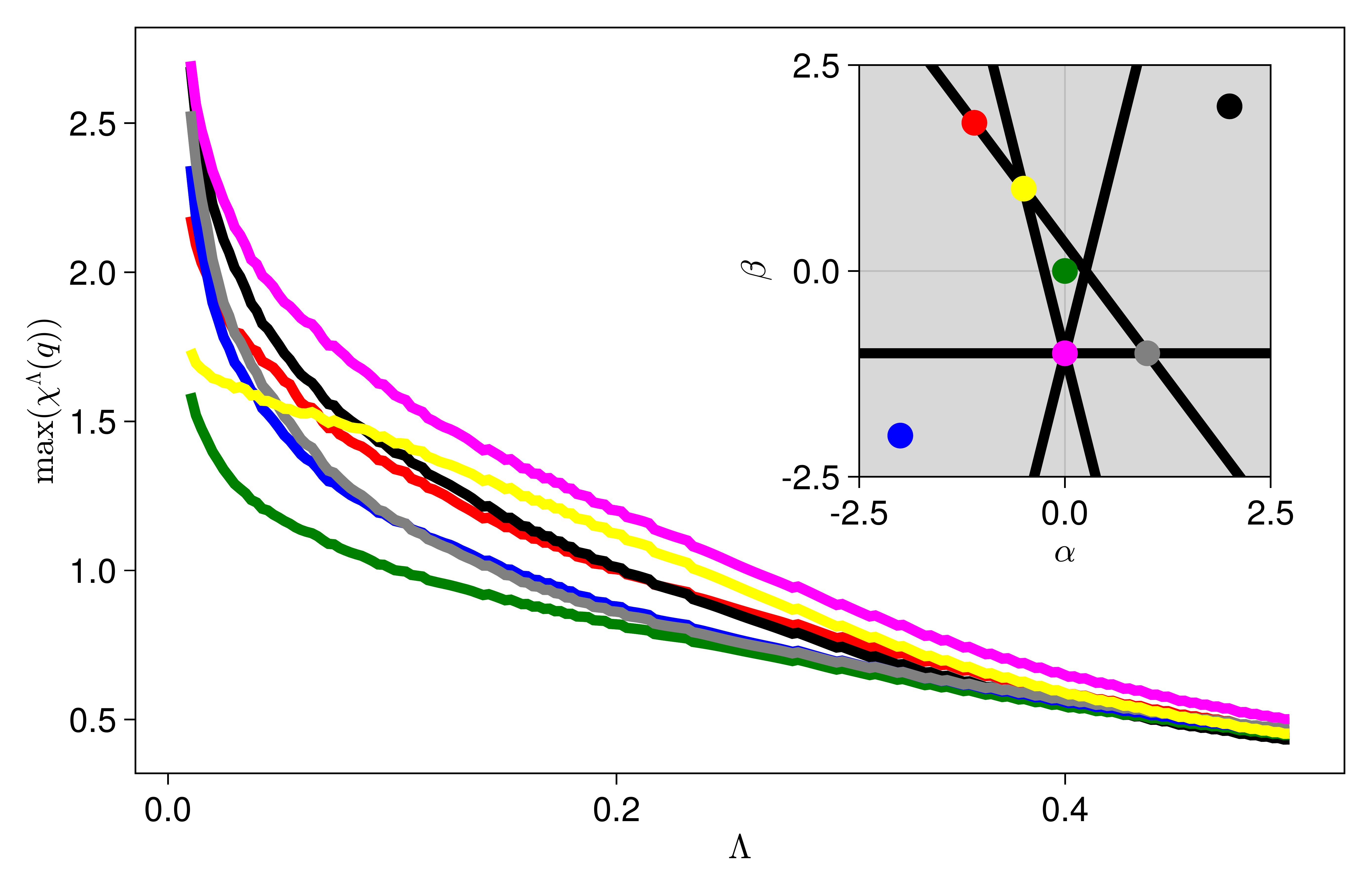}
    \caption{\textbf{Selected PFFRG flows:} Flows of the maximal zero-frequency Matsubara susceptibility obtained from PFFRG for an exemplary selection of points along boundaries and the bulk of phases in the phase diagram. No feature or flow breakdown is observed, indicating absence of magnetic ordering. }
    \label{fig:SelectedFlows}
\end{figure}

\section*{Methodological Details}

\begin{figure}[b]
    \centering
    \includegraphics[width = \linewidth]{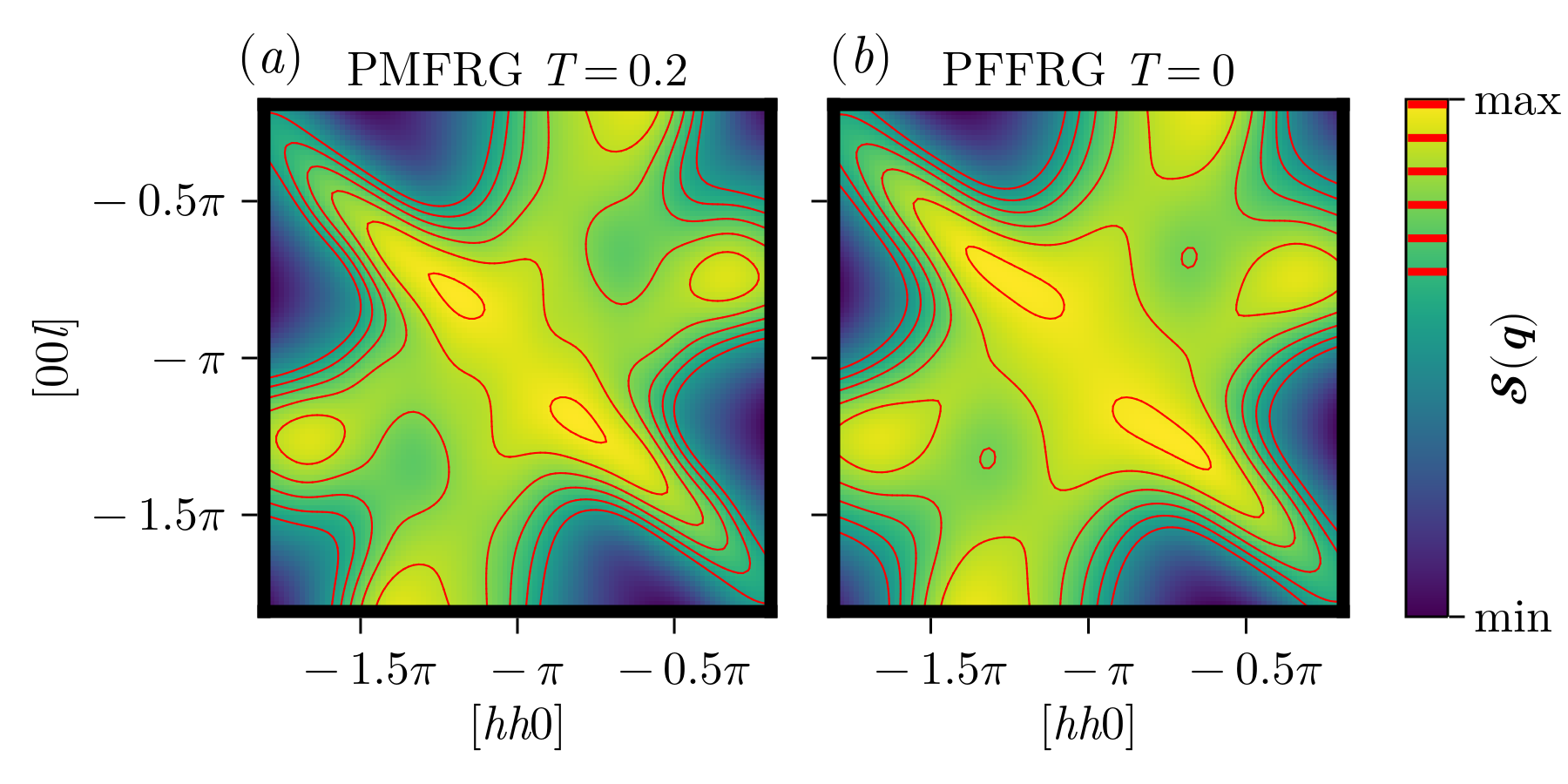}
    \caption{\textbf{Spin structure factor from PMFRG at $T=0.2$ (a) and PFFRG at $T=0$ (b)} for the multifold pinch point at $\alpha = -\frac{11}{10}$, $\beta = \frac{9}{5}$. The color scale is normalized to the extrema of the susceptibility and the relative location of the contours is indicated.}
    \label{fig:PMPFAgreement}
\end{figure}

For this work, standard PFFRG and PMFRG implementations were used, further methodological details are found in \cite{ReutherOrig,Niggemann2021,Niggemann2022}. 
In both cases, a large set of ordinary differential equations are computed for  vertices with up to $N = 64$ positive Matsubara frequencies (well beyond convergence) and correlations up to $L = 10$ nearest neighbor bonds using an adaptive Runge-Kutta scheme. For the largest temperatures $T>1$, a smaller number of $N=40$ was used instead, as convergence is reached even more rapidly.
We note that while the FRG's maximal length of spin correlations $L$ is a numerical requirement, the shape of the structure factor converges already before $L=10$. As a result, the finite width of pinch points, (and, accordingly, the deviation from dipolar correlations) is not a numerical artifact but due to a physically finite correlation length. In such cases, where the physical correlation length is smaller than the numerical length $L$, results obtained via FRG correspond to the thermodynamic limit.    
The standard output of these calculations is the magnetic susceptibility in Matsubara frequency space
\begin{equation}
    \chi_{ij}(\imath \nu_n) = \int_0^\beta d\tau \langle S^z_i(\tau) S^z_j(0) \rangle e^{\imath \nu_n \tau}.\label{eq:chi}
\end{equation}
In PFFRG, magnetic order is indicated by a sharp feature in the Fourier transform of the $\nu=0$ component of \cref{eq:chi} for the flow of the order-defining momentum $\bmq$. The absence of such order for a variety of points in the phase diagram is demonstrated in \cref{fig:SelectedFlows}. 
In all other results discussed in this work, we instead consider the equal time structure factor which is obtained as $\mathcal{S}(\bmq) \equiv \langle S^z(-\bmq) S^z(\bmq) \rangle = \frac{T}{N_\textrm{sites}} \sum_n  \sum_{ij} \chi_{ij}(i\nu_n) e^{\imath \bm{q} \bm{R}_{ij}}$, where $\bm{R}_{ij}$ is the displacement vector between two sites for better comparison with the large-$N$ approximation.
Despite the methodological differences, we observe good qualitative agreement between the structure factor from PMFRG at very low temperature and PFFRG as demonstrated in \cref{fig:PMPFAgreement} for the $Q=-7$ multifold pinch point.

In the PMFRG's underlying $SO(3)$ Majorana representation, a local constant of motion $\theta_i = -2i \eta^x_i \eta^y_i \eta^z_i,\ \theta^2_i = 1/2$ can be used to derive an exact Ward identity relating the two-point Green's function to the local four-point vertex as $\langle \eta^z_i(\tau)\eta^z_i(0) \rangle = 2 \langle \eta^x_i(\tau)\eta^y_i(\tau) \eta^x_i(0)\eta^y_i(0) \rangle$. In a finite truncation of the flow equations, this identity is only fulfilled approximately and can therefore be used as a rigourous accuracy check. In this work, the relative violation of this Ward identity is remains below $\sim 10 \%$ for all temperatures down to $T=0.2$.

\end{document}